\documentclass[twoside,twocolumn,american,sort&compress]{revtex4-1}
\usepackage[T1]{fontenc}
\usepackage[latin9]{inputenc}
\usepackage{float}
\usepackage{amsmath}
\usepackage{graphicx}

\makeatletter
\usepackage{color}
\usepackage{hyperref}
\hypersetup{
     colorlinks   = true,
     citecolor    = blue
}

\makeatother

\usepackage{babel}
\begin{document}

\title{Analytical study of the edge states in the bosonic Haldane model}

\author{Pierre A. Pantale{\'o}n}

\address{Theoretical Physics Division, School of Physics and Astronomy, University
of Manchester, Manchester M13 9PL, United Kingdom}

\author{Y. Xian}

\address{Theoretical Physics Division, School of Physics and Astronomy, University
of Manchester, Manchester M13 9PL, United Kingdom}
\begin{abstract}
\noindent {\normalsize{}We investigate the properties of magnon edge
states in a ferromagnetic honeycomb spin lattice with a Dzialozinskii-Moriya
interaction (DMI). We derive analytical expressions for the energy
spectra and wavefunctions of the edge states localized on the boundaries.
By introducing an external on-site potential at the outermost sites,
we show that the bosonic band structure is similar to that of the
fermionic graphene. We investigate the region in the momentum space
where the bosonic edge states are well defined and we analyze the
width of the edge state and their dependence with the DMI strength.
Our findings extend the predictions using topological arguments and
they allow size-dependent confirmation from possible experiments. }{\normalsize \par}
\end{abstract}
\maketitle

\section{Introduction}

Perhaps the most important and intriguing aspect of the topological
insulators is the presence of edge states. A well known example is
the Kane-Mele model in graphene \cite{Kane2005}, where the spin-orbit
coupling (SOC) causes a transition from a semi-metal to a quantum
spin Hall insulator. Accompanying the transition is the appearing
of gapless edge states with distinct properties from the bulk energy
band \cite{Gusynin2008,Wakabayashi2010a,Konschuh2010}. In a fermionic
model, the edge states are spin-filtered, since the electrons with
opposite spins propagate in opposite directions with robustness against
external perturbations \cite{Yao2009,Konig2013}.

The edge states have also been studied in magnetic insulators \cite{Fujimoto2009},
where the spin moments are carried by magnons. The magnon Hall effect
was observed in a collinear ferromagnetic insulator $\textrm{Lu}_{2}V_{2}O_{7}$
with pyrochlore structure \cite{Onose2010}, in the K{\'a}gome ferromagnetic
lattice \cite{Chisnell2015}, and have also been studied in the Lieb
lattice \cite{Cao2015} and honeycomb ferromagnetic lattice \cite{Owerre2016d}.

Until recently, most investigations of the edge states have been based
on fermionic models. Now similar study has been extended to the bosonic
models. For example, it has been recently shown that the bosonic equivalent
for the Kane-Mele-Haldane model is a ferromagnetic Heisenberg Hamiltonian
with the Dzialozinskii-Moriya interaction. The thermal Hall effect
\cite{Owerre2016c} and spin Nernst effect \cite{Kim2016a} have been
predicted for this magnetic system. So far, the study of the magnon
edge states is mainly based on topological arguments and clearly the
detailed properties of the edge states will be required if the potential
of magnonics is to be realized \cite{Kruglyak2010}. In particular,
systematic investigation of the spin-density profile (or magnon density)
of the edge magnon and their dependence with the DMI strength and
external on-site potentials may be useful in manufacturing small-sized
devices based on such magnets.

Although a direct experimental demonstration of the magnon edge modes
in magnetic systems is still absent, it is interesting to ask whether
the similar edge properties in the fermionic systems are also exhibited
in the bosonic counterparts. Particularly, with regard to the fact
that the strength of DMI in some magnets is greater than that of SOC
in graphene by five orders of magnitude \cite{Onose2010,Chisnell2015}.
With such motivations in this paper we analyze a ferromagnetic honeycomb
lattice with DMI. After introducing an on-site potential at the outermost
sites \cite{Sakaguchi2016,Guo2016} we derive an analytical form for
the wavefunctions and energy spectrum for the edge states. We investigate
the conditions in the momentum space where the edge states are well
defined and we write the edge-state width in terms of the DMI strength.

This paper is organized as follows: In Sec. II we introduce the model
Hamiltonian with a next-nearest neighbor DMI which opens a gap in
the energy spectra. In Sec. III we adapt the analytical approach for
the edge state developed by \textit{Wang. et. al.} \cite{Wang2009}
and \textit{Doh. et. al} \cite{Doh2013,Doh2014} to a bosonic Haldane
model. In order to cover the region where most study has been focusing
using topological approach, we introduced an on-site potential at
the outermost sites. In Sec. IV, the edge states and their energy
spectra for zero and non-zero DMI are investigated. We also analyze
the range in the momentum space in which the edge states are well
defined and we express the width of the edge state in terms of the
DMI strength. We conclude with a summary in Sec. V. A brief description
to obtain the edge states and the energy spectrum without DMI is presented
in the appendix.

\section{Model Hamiltonian}

\begin{figure}[H]
\begin{centering}
\includegraphics[scale=0.3]{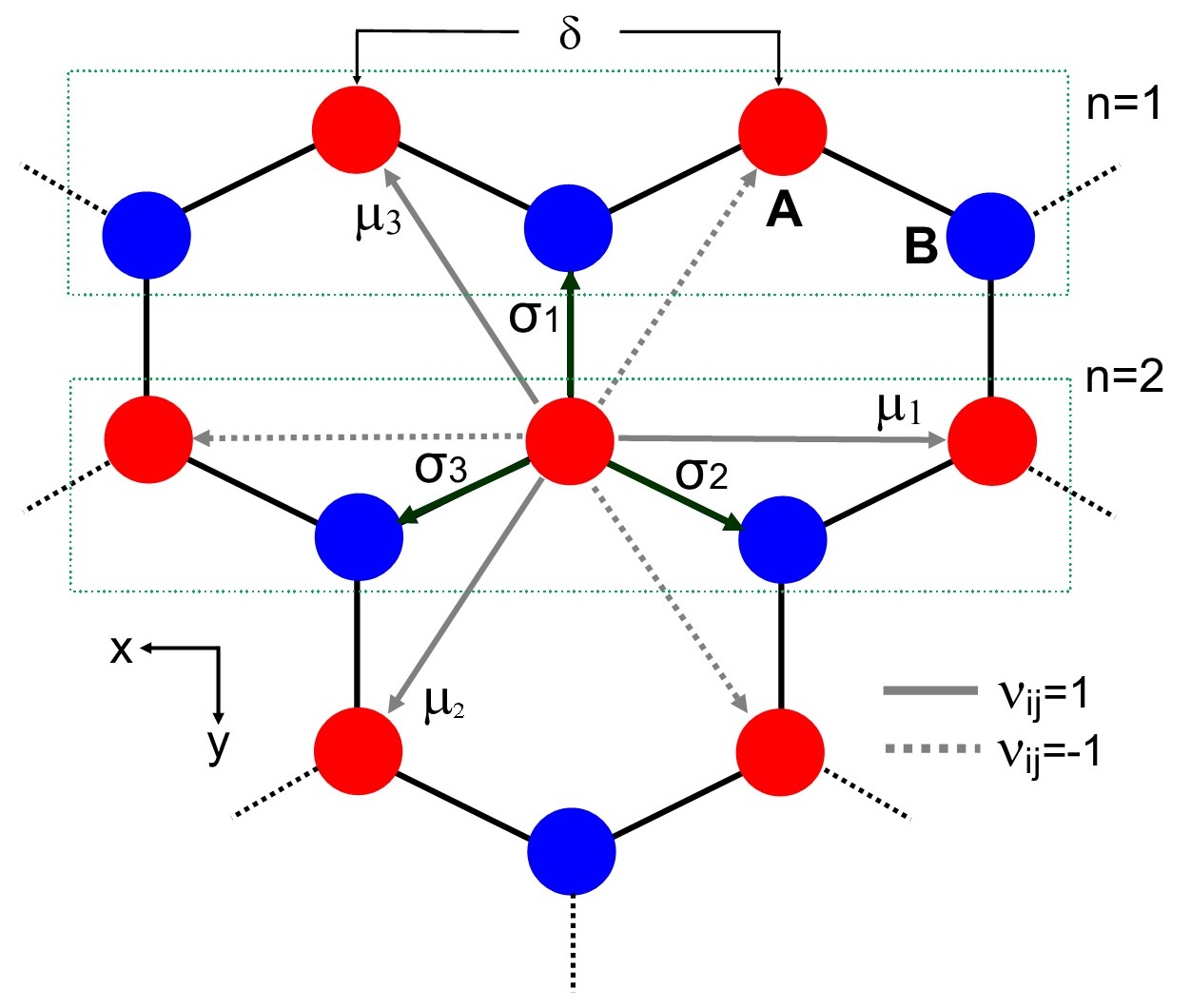}
\par\end{centering}
\caption{{\footnotesize{}(Color on-line) Schematic of a honeycomb lattice with
zig-zag edges. $\sigma,\,\mu$ are the nearest-neighbors and the next-nearest
neighbor index vectors, respectively. $\nu_{ij}=\pm1$ is the orientation
dependent coefficient and is positive (negative) if the electron makes
a left (right) turn to get the second site. The external on-site potential
$\delta=1$ is introduced at the outermost sites. }{\small{}\label{fig: Figure1}}}

\end{figure}

We consider a ferromagnetic Heisenberg Hamiltonian with DMI on a honeycomb
lattice. The DMI vanishes for the nearest-neighbors (NN) but has non-zero
component along the $z-$direction for the next-nearest-neighbors
(NNN) \cite{Moriya1960}. Assuming $\overrightarrow{D}=D\hat{\mathbf{z}}$,
the corresponding model Hamiltonian for $J>0$ is
\begin{equation}
H=-J\sum_{\left\langle i,j\right\rangle }\mathbf{S}_{i}\cdot\mathbf{S}_{j}+D\sum_{\left\langle \left\langle i,j\right\rangle \right\rangle }\nu_{ij}\hat{\mathbf{z}}\cdot\left(\mathbf{S}_{i}\times\mathbf{S}_{j}\right),\label{eq:HFull}
\end{equation}
where $\left\langle i,j\right\rangle $ indicates the NN coupling,
$\left\langle \left\langle i,j\right\rangle \right\rangle $ is the
NNN coupling, in analogy with the Kane-Mele model, and $\nu_{ij}=\pm1$,
depending on the orientation of the two sites \cite{Kane2005}. The
energy spectrum of Eq. (\ref{eq:HFull}) is given by
\begin{equation}
\varepsilon=3JS\left(1\pm\sqrt{h_{d}^{2}+\left|\gamma_{k}\right|^{2}}\right),\label{eq: BulkEner}
\end{equation}
in the linear spin-wave approximation for the infinite system. Here,
$\gamma_{k}=\frac{1}{3}\sum_{\sigma}\exp\left(k\cdot\sigma\right)$
is the honeycomb complex structure factor and $h_{d}=\frac{2}{3}\frac{D}{J}\sum_{\mu}\sin\left(k\cdot\mu\right)$
is the contribution due to the DMI. The index vectors $\sigma$ and
$\mu$ run over the three NN and NNN with positive hopping term, as
shown in Fig. (\ref{fig: Figure1}). The lower energy band becomes
negative if $\frac{D}{J}>\frac{1}{\sqrt{3}}$, which is a signature
of a phase transition in the system \cite{Kim2016a}. Here, we only
consider the case when $D<\frac{J}{\sqrt{3}}$ with the ferromagnetic
ground state.

The system of Eq. (\ref{eq:HFull}) has a non-trivial band topology.
The DMI induces a non-trivial gap of $\frac{6\sqrt{3}D}{J}$ at the
two Dirac points $\mathbf{K}$ and $\mathbf{K^{\prime}}$ in the spin-wave
spectra and edge states emerge with gapless energy \cite{Owerre2016d}.
Motivated by such predictions, in the following sections, we derive
analytical expressions for the edge states and explicit forms for
the energy spectrum for the semi-infinite system.

\section{Edge States and Boundary Conditions}

\subsection{Hamiltonian matrix elements}

Starting from a Ne{\'e}l ordering state for a bipartite lattice and
by the standard Holstein-Primakoff transformations in the linear spin-wave
theory, we can write the following effective bosonic Hamiltonian for
a ferromagnetic lattice of Eq. (\ref{eq:HFull}) as,
\begin{eqnarray}
H & = & -JS\sum_{i,\sigma}\left(a_{i}b_{i+\sigma}^{\dagger}+a_{i}^{\dagger}b_{i+\sigma}-a_{i}^{\dagger}a_{i}-b_{i+\sigma}^{\dagger}b_{i+\sigma}\right)\nonumber \\
 &  & +iDS\sum_{i,\mu}\nu_{i,i+\mu}\left(a_{i}a_{i+\mu}^{\dagger}-a_{i}^{\dagger}a_{i+\mu}\right)\label{eq: HaldaneHamil}\\
 &  & +iDS\sum_{j,\mu}\nu_{j,j+\mu}\left(b_{j}b_{j+\mu}^{\dagger}-b_{j}^{\dagger}b_{j+\mu}\right),\nonumber
\end{eqnarray}
where the index $i\left(j\right)$ denotes the $A\,(B)$ sublattice
sites, $\sigma$ and $\mu$ are the NN and NNN vector, respectively,
as shown in Fig. (\ref{fig: Figure1}). This Hamiltonian is the bosonic
equivalent to the Haldane model \cite{Haldane1988}. To investigate
the edge state, we assume a zigzag boundary along the $x$ direction
and semi-infinite in the $y$ direction. If $k$ is the momentum in
the $x$ direction, the Hamiltonian (\ref{eq: HaldaneHamil}) can
be written as,
\begin{equation}
H=JS\sum_{k}\Psi^{\dagger}M\Psi,\label{eq: HolePartHamil}
\end{equation}
where $\Psi_{k}^{\dagger}=\left[\Psi_{k,A}^{\dagger},\,\Psi_{k,B}^{\dagger}\right]$
is a semi-infinite $2$-component spinor. The matrix elements of $M$
are semi-infinite matrices: $M_{11}=(2+\delta)+(1-\delta)T^{\dagger}T+J_{3}-J_{4}\left(T+T^{\dagger}\right)$,
$M_{12}=-\left(J_{1}+J_{2}T^{\dagger}\right)$, $M_{21}=M_{12}^{\dagger}$
and $M_{22}=(2+\delta)+(1-\delta)TT^{\dagger}-J_{3}+J_{4}\left(T+T^{\dagger}\right)$,
where $J_{1}=2\cos\left(\frac{\sqrt{3}}{2}k\right)$, $J_{2}=1$,
$J_{3}=2\,D^{\prime}\,\sin\left(\sqrt{3}\,k\right)$, $J_{4}=2\,D^{\prime}\,\sin\left(\frac{\sqrt{3}}{2}k\right)$,
$D^{\prime}=D/J$, and $T$ is a semi-infinite displacement matrix
as defined in Ref. \cite{You2008a}. Here, we introduced an external
on-site potential, $\delta$, at the outermost sites. In a ferromagnetic
lattice the intrinsic on-site potential is given by the number of
nearest-neighbors, along the zig-zag edge there is a missing bond
and the edge on-site energy is lower than in the bulk. By introducing
an external on-site potential the edge state can be modified \cite{Yao2009,Sakaguchi2016,Semenoff2008}.
In this work, we will focus on the ``symmetric'' case with $\delta=1$,
as in Refs. \cite{Owerre2016d,Owerre2016c,Kim2016a}, where the on-site
potential along the edge is the same as in the bulk. We will consider
other cases with different values of $\delta$ elsewhere.

\subsection{Self-consistent equation for the edge modes}

From the Hamiltonian (\ref{eq: HolePartHamil}) and with the explicit
form of the matrix elements, we can construct the coupled Harper equations
\cite{Harper1955,Hofstadter1976} for the edge state. Using the assumption
that the edge states are exponentially decaying from the boundary,
we consider the following anzats \cite{Konig2008,Wang2009} for the
eigenstates of $M$ in Eq. (\ref{eq: HolePartHamil}),
\begin{equation}
\psi_{k}\left(n\right)=z^{n}\phi_{k}.\label{eq: MainAnzats}
\end{equation}
In the above wavefunction $\phi_{k}$ is a $2-$component vector,
$z$ is a complex number with magnitude less than the unity and $n=1,\,2,\,3\,...$,
are the lattice coordinates in the $y$ direction as shown in Fig.
(\ref{fig: Figure1}). The form $z^{n}$ in the above equation is
a finite-size Fourier transform along the $y$-direction and is related
to the eigenvalues of a finite-size displacement matrix \cite{You2008a}.
Therefore, the effective Hamiltonian for the edge state can now be
written as, {\small{}
\begin{equation}
H=\left[\begin{array}{cc}
3+J_{3}-J_{4}\left(z+z^{-1}\right) & -\left(J_{1}+J_{2}z^{-1}\right)\\
-\left(J_{1}+J_{2}z\right) & 3-J_{3}+J_{4}\left(z+z^{-1}\right)
\end{array}\right].\label{eq: EdgeHamil}
\end{equation}
}{\small \par}

The characteristic equation for the eigenvalue $\varepsilon$ is obtained
from the condition, $\left|H\left(z\right)-\varepsilon\right|=0$,
or,
\begin{equation}
a\left(z+z^{-1}\right)^{2}+b\left(z+z^{-1}\right)+c=0,\label{eq: zCharact}
\end{equation}
where, $a=-J_{4}^{2}$, $b=2J_{3}J_{4}-J_{1}J_{2}$, $c=\left(\varepsilon-3\right)^{2}-J_{1}^{2}-J_{2}^{2}-J_{3}^{2}$.
We solve for $z+z^{-1}$ to obtain four solutions $z_{1}^{\pm}$ and
$z_{2}^{\pm}$ in terms of the momentum $k$ and energy $\varepsilon$.
Two of the solutions with $\left|z\right|<1$ describe the state at
the upper edge, while the other two with $\left|z\right|>1$ describe
the state for the lower edge, if any. The eigenfunction of Eq. (\ref{eq: HolePartHamil})
satisfying $\underset{n\rightarrow\infty}{lim}\,\psi_{k}\left(n\right)=0$
may now in general be written as,
\begin{equation}
\psi_{k}\left(n\right)=u_{1}z_{1}^{n}\phi_{1}+u_{2}z_{2}^{n}\phi_{2},\label{eq: LinCom}
\end{equation}
where $u_{1(2)}$ are normalization constants and the eigenvector
of the Hamiltonian, Eq. (\ref{eq: EdgeHamil}), corresponding to the
solution $z_{r}$ $(r=1,2)$, is given by
\begin{equation}
\phi_{r}=\left[\begin{array}{c}
J_{1}+J_{2}z_{r}^{-1}\\
3-\varepsilon+J_{3}-J_{4}\left(z_{r}+z_{r}^{-1}\right)
\end{array}\right].\label{eq: EigenUpper}
\end{equation}

If an opposite boundary exists, the semi-infinite matrices in Eq.
(\ref{eq: HolePartHamil}) can be truncated at some size $N$ and
the edge states of the system will become a linear combination of
the solutions at each boundary \cite{Wakabayashi2010a}. As $N$ increases
over some value, the edge states become independent of the size as
expected \cite{Hatsugai1993,Hatsugai1993a}, allowing us to obtain
the solutions at each edge independently. For a large $N$ and for
the opposite boundary, the edge state can be obtained if we perform
the substitution $z^{n}\rightarrow z^{N-n+1}$ in the Eq. (\ref{eq: LinCom}).
Furthermore, we have to choose the eigenvector whose $A-$sublattice
component goes to zero when $D\rightarrow0$ (the details of the solution
without DMI are presented in the appendix). From the Hamiltonian (\ref{eq: EdgeHamil}),
the eigenvector, $\phi_{r}^{\prime}$, for the opposite edge, has
the form,
\begin{equation}
\phi_{r}^{\prime}=\left[\begin{array}{c}
3-\varepsilon-J_{3}+J_{4}\left(z_{r}+z_{r}^{-1}\right)\\
J_{1}+J_{2}z_{r}
\end{array}\right].\label{eq: EigenLower}
\end{equation}

For a zig-zag edge, the boundary conditions are satisfied by setting:
\begin{equation}
\psi_{k}^{\prime}\left(N+1\right)=\psi_{k}\left(0\right)=0.\label{eq: EdgeCond}
\end{equation}
As in the Kane-Mele model, the eigenvectors for each solution are
linearly dependent \cite{Doh2014}. Then, from the Eq. (\ref{eq: LinCom})
and the edge condition, Eq. (\ref{eq: EdgeCond}), the non-trivial
solution is given by setting the determinant $\left|\phi_{1}\,\phi_{2}\right|=0$
or $\left|\phi_{1}^{\prime}\,\phi_{2}^{\prime}\right|=0$ for the
other edge, providing us the following additional relation for the
edge energy dispersion,
\begin{equation}
\varepsilon-3-2J_{3}+J_{4}\left(z_{1}+z_{2}\right)+J_{3}z_{1}z_{2}=0,\label{eq: AddEdge}
\end{equation}
with eigenstates
\begin{equation}
\psi_{k}\left(n\right)=u_{1}\left(z_{1}^{n}-z_{2}^{n}\right)\phi_{1},\label{eq:EdgeCompactAll-1}
\end{equation}
for the upper edge. As we mentioned before, for the lower edge at
large $N$, the eigenstate and energy spectrum can be obtained by
the substitutions $z^{n}\rightarrow z^{N-n+1}$ in the Eq. (\ref{eq:EdgeCompactAll-1})
and $z\rightarrow z^{-1}$ in the Eq. (\ref{eq: AddEdge}).
\begin{figure}[H]
\centering{}\includegraphics[scale=0.35]{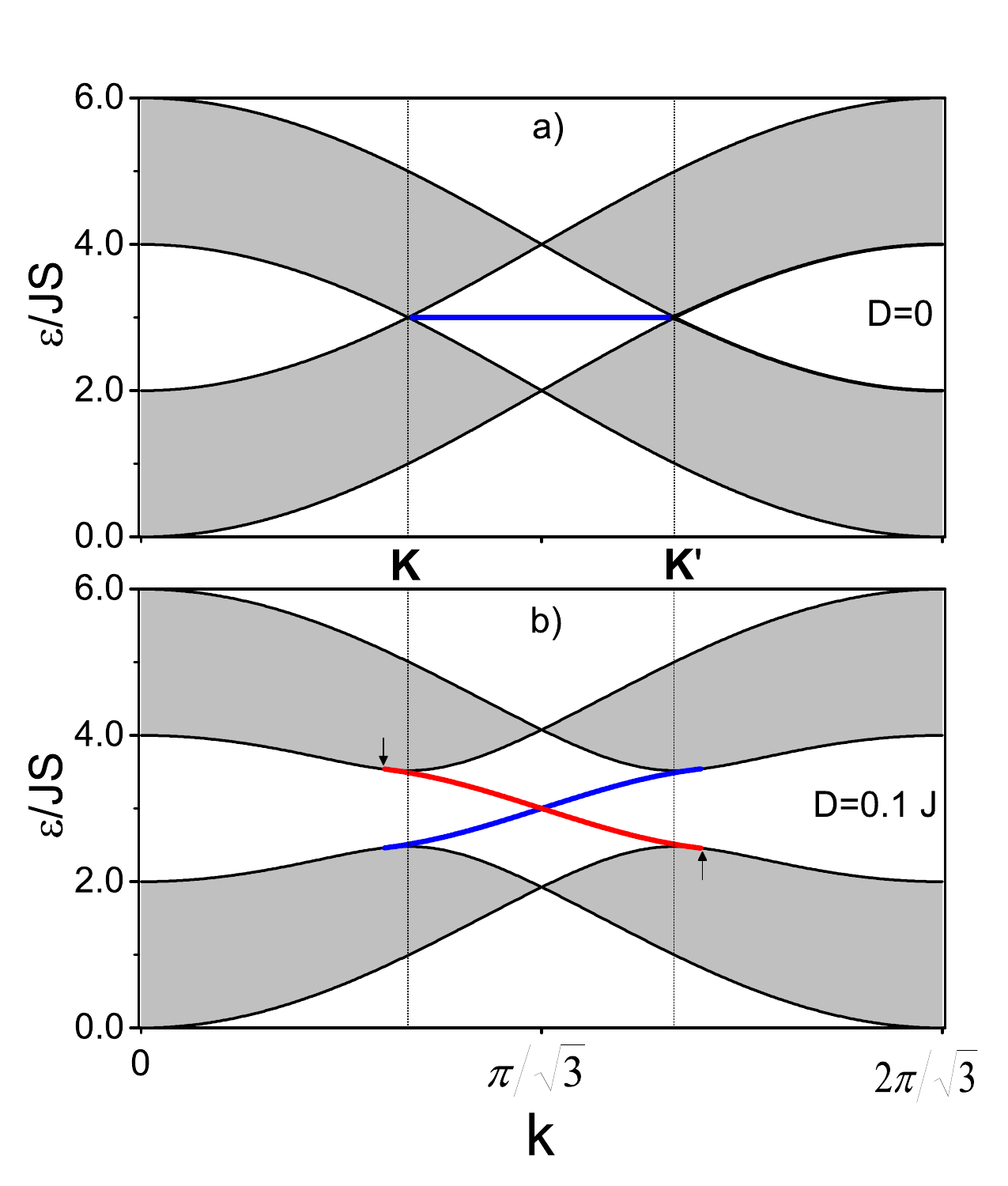}\caption{{\footnotesize{}(Color online) Energy spectrum for a honeycomb ferromagnetic
lattice with DMI and external on-site potential $\delta=1$ at the
outermost sites. a) For $D=0$, the horizontal line at $\varepsilon=3\,JS$
is the flat spectrum for the edge state. b) For $D=0.1\,J$, the two
bands crossing the gap near the Dirac points are the chiral edge states
for the upper (red) and lower (blue) edges. The arrows show the critical
points at $k_{l}$ before $\mathbf{K}$ and $k_{r}$ after $\mathbf{K^{\prime}}$
(see text)}{\small{}.\label{fig: Figure2}}}

\end{figure}

\section{Results and discussions}

\subsection{Energy Spectrum}

Equations (\ref{eq: zCharact}) and (\ref{eq: AddEdge}) can be solved
numerically for the energy and the wavefunction of the edge state.
We first discuss the case of $D=0$. As shown in Fig. \ref{fig: Figure2}(a),
the bulk energy spectra is always degenerate at the $\mathbf{K}$
and $\mathbf{K^{\prime}}$ points, independently of the boundary conditions.
However, if we consider a semi-infinite lattice with zig-zag edges,
a flat band appears (see appendix for further details). Such flat
band is the bosonic analogous to the edge state in zig-zag graphene
\cite{Fujita1996,Nakada1996} and the corresponding edge state can
be described in a similar way as follows: the bulk bands for a ferromagnetic
lattice are symmetric around $\varepsilon=3$, hence, if a state exists
along the boundary, the sum of the components of the wavefunction
over the nearest-neighbors sites must be equal to the on-site energy.
Such equality is only fulfilled if we include an external on-site
energy, in particular for $\delta=1$ we obtain an edge state localized
at the $A-$sublattice. From the Eq. (\ref{eq: UpEigenfree}), we
can write the wavefunction as,
\begin{equation}
\psi_{k}\left(n\right)\propto\left(-J_{1}\right)^{n-1},\label{eq:WaveFreeD0}
\end{equation}
where $J_{1}$ is a function of $k$ as we defined before. Hence,
the spin-density of the edge magnon is proportional to $J_{1}^{2(n-1)}$
in the region $\frac{2\pi}{3\sqrt{3}}\leq k\leq\frac{4\pi}{3\sqrt{3}}$
(between the $\mathbf{K}$ and $\mathbf{K^{\prime}}$ points) where
the flat band exists. At the point $k_{0}=\frac{\pi}{\sqrt{3}}$ the
edge state is completely localized at the single line of boundary
sites and gradually extends inside to the inner sites as we move to
the Dirac Points where it merges into the bulk. For the opposite edge,
with the same argument, an edge state exists and is localized at the
$B-$sublattice.

We next consider the case of non-zero DMI. The energy bands are obtained
by solving the Eq. (\ref{eq: AddEdge}) with the solutions given by
the Eq. (\ref{eq: zCharact}), and the edge wavefunction by the Eq.
(\ref{eq:EdgeCompactAll-1}). In the Fig. \ref{fig: Figure2}(b) we
show the energy bands for a DMI strength of $D=0.1\,J$. The shaded
regions correspond to the bulk spectra, while the two bands which
transverse the gap, connecting the upper and lower bulk bands, are
the spectra of the edge states. By the analysis of the Eq. (\ref{eq: AddEdge})
or the wavefunctions of Eq. (\ref{eq:EdgeCompactAll-1}), we find
that there is only one edge state at each boundary. The edge band
at the upper boundary has negative slope while the edge band at the
opposite edge has positive slope. Hence, as predicted in Ref. \cite{Owerre2016d},
the magnons are moving to the left at one of the edges, while they
are moving to the right in the other one. On the other hand, at the
point $k_{0}=\frac{\pi}{\sqrt{3}}$ and $\varepsilon=3$, the edge
state for the upper edge takes a simple form,
\begin{equation}
\psi_{k_{0}}\left(n\right)\propto\left[1-\left(-1\right)^{n}\right]\left[\begin{array}{c}
z_{0}^{-1}\\
i
\end{array}\right]z_{0}^{n},\label{eq:EdgeDko}
\end{equation}
where $z_{0}=\frac{i}{4D^{\prime}}\left(-1+\sqrt{1+16D^{\prime2}}\right)$.
The spin-density of the edge magnon is therefore proportional to $\left|z_{0}\right|^{2(n-1)}$
and $\left|z_{0}\right|^{2n}$ for the $A$ and $B$ sublattice, respectively,
decaying exponentially. As we move from $k_{0}$ to near the Dirac
points $\mathbf{K}$ or $\mathbf{K^{\prime}}$, the edge state becomes
delocalized and merged with that of the bulk state.

\subsection{Allowed range for the edge states}

\begin{figure}[H]
\begin{centering}
\includegraphics[scale=0.35]{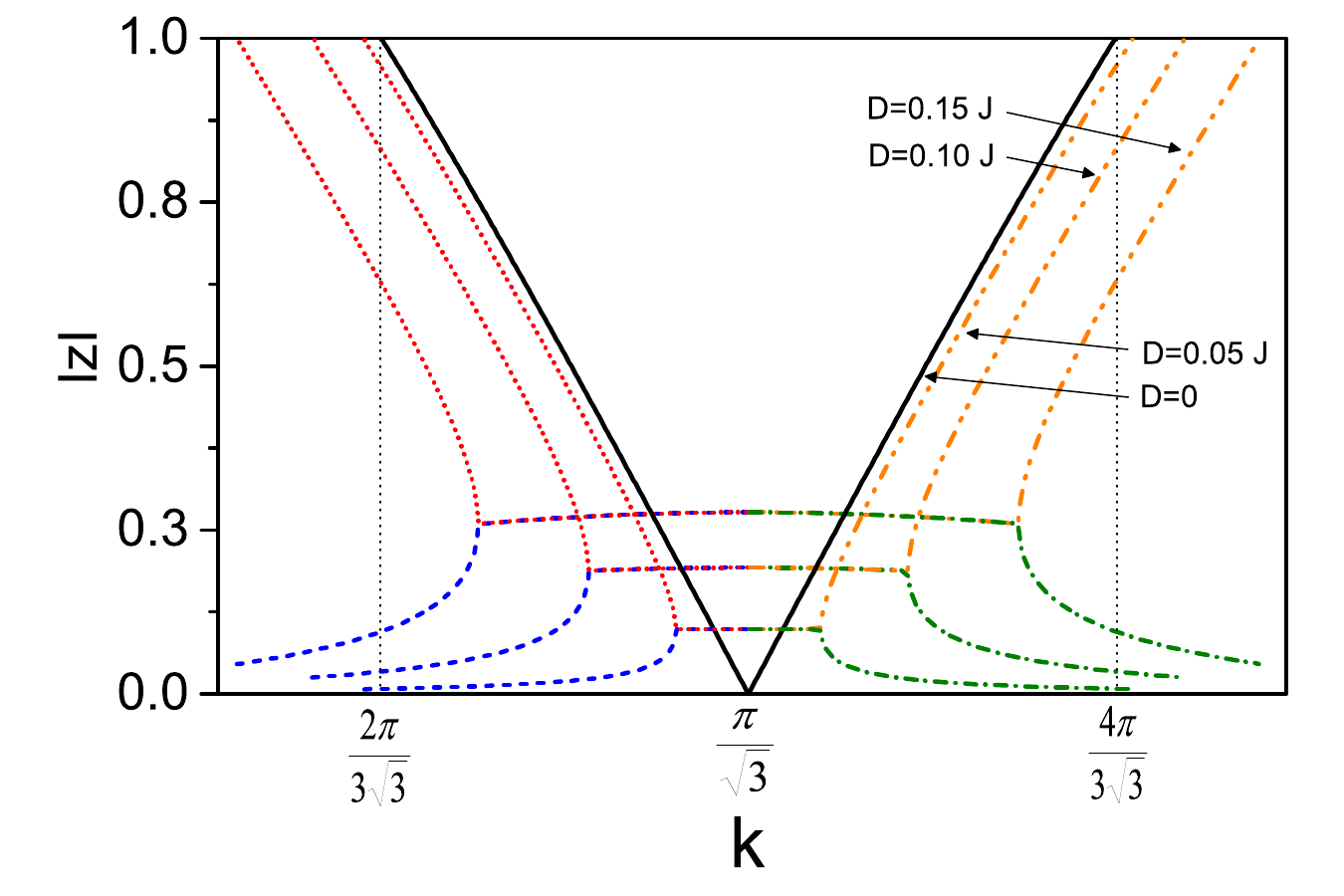}
\par\end{centering}
\caption{{\footnotesize{}(Color on-line) $\left|z\right|$ vs $k$ at different
values of $D$. The solutions correspond to $\left|z_{1}^{+}\right|$
(dashed, blue), $\left|z_{2}^{+}\right|$ (dot, red), $\left|z_{2}^{-}\right|$
(dot-dashed, green) and $\left|z_{1}^{-}\right|$ (dot-dot-dashed,
orange). The continuous line (black) corresponds to $z=-J_{1}$ for
$D=0$.}{\small{} \label{fig: Figure3}}}

\end{figure}

We notice that the restriction $\left|z_{1,2}\right|<1$ in the Eq.
(\ref{eq: AddEdge}) and (\ref{eq:EdgeCompactAll-1}) allow us to
investigate two important properties of the edge states: the region
in the momentum space where they are well defined and their confinement
to the physical boundary. In general, a particular edge state has
two critical points defined as $k_{l}$ and $k_{r}$ located before
and after the two Dirac $\mathbf{K}$ and $\mathbf{K}^{\prime}$ points,
respectively, as indicated in the Fig. \ref{fig: Figure2}(b). For
the upper edge, the eigenfunctions (and also the energy spectrum)
depends on $z_{1}^{+}$ and $z_{2}^{+}$ in the (left) interval $k\in\left(k_{l},\,k_{0}\right)$
and on $z_{1}^{-}$ and $z_{2}^{-}$ in the (right) interval $k\in\left(k_{0},k_{r}\right)$.
The critical points are determined when the modulus of one $z$ in
Eq. (\ref{eq: AddEdge}) reach the unity and the edge state becomes
indistinguishable from the bulk. In the Fig. (\ref{fig: Figure3})
the plot of $\left|z\right|$ versus $k$ is shown for different values
of $D$ in the interval where the edge state is well defined. For
the edge state at the upper edge, the critical point near $\mathbf{K}$
is given when the modulus $\left|z_{2}^{+}\right|$, reach the unity,
whereas that in the right interval (near $\mathbf{K}^{\prime}$),
the critical point is given when $\left|z_{1}^{-}\right|=1$. In particular
for $D=0.1\,J$, $k_{l}\approx1.09$ and $k_{r}\approx2.53$ {[}See.
Fig. \ref{fig: Figure2}(b){]}. Only when $D=0$, these critical points
correspond to the Dirac points.

If we increase the value of $D$, the critical points move further
away from the Dirac points as shown in Fig. (\ref{fig: Figure3}).
This suggests that the edge state merges into the bulk but remains
in (or near) its boundaries. Such critical points are important since
they define the range where the edge state is well defined and they
also determine the critical region where $z$ is transformed from
real to a complex with unitary modulus. The $z$ solutions with unitary
modulus, which obey the bulk and boundary conditions, are sensitive
to the values of DMI and the details of edge modifications.

\subsection{Width of the bosonic edge state}

\begin{figure}[H]
\begin{centering}
\includegraphics[scale=0.36]{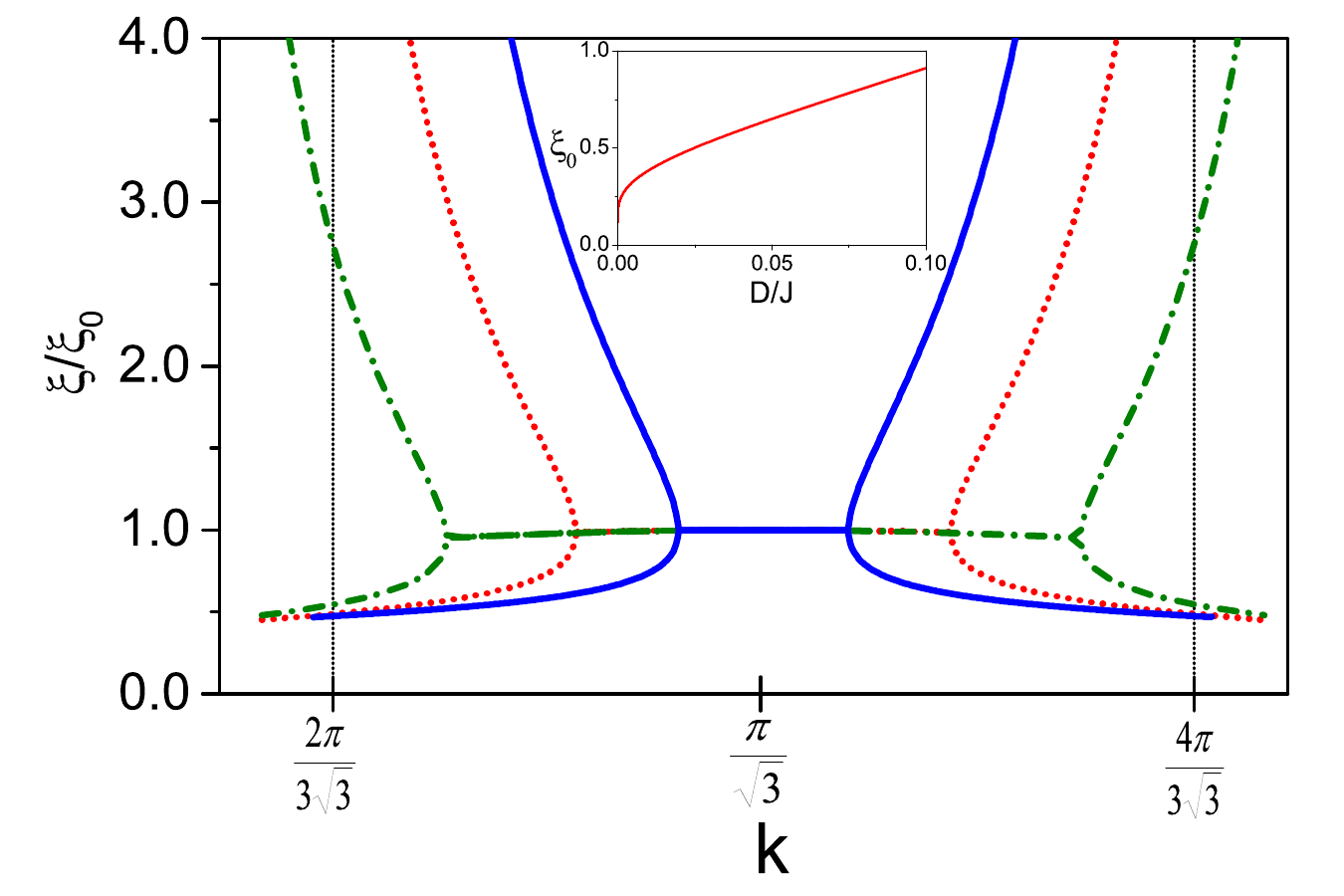}
\par\end{centering}
\caption{{\footnotesize{}(Color on-line) Edge state width $\xi$ vs $k$ at
different values of $D$. The DMI values correspond to: $D=0.05\,J$
(continuous, blue), $D=0.10\,J$ (dotted, red) and $D=0.15\,J$ (dash-dotted,
green). Inset: Edge state width at $k_{0}=\frac{\pi}{\sqrt{3}}$ as
a function of $D/J$.}{\small{} \label{fig: Figure4}}}

\end{figure}

In the previous section, we have investigated the critical points
and their dependence with the solutions $z_{2}^{+}$ and $z_{1}^{-}$
in their corresponding intervals. In this section, we will examine
the width of the edge state and their relation with the remaining
solutions, $z_{1}^{+}$ and $z_{2}^{-}$. As described before, the
spin-density of the edge magnon decreases as we move away from the
boundary, and this happens with some characteristic length scale.
Following the Ref. \cite{Doh2013} for a semi-infinite zig-zag graphene,
the width or characteristic length scale is given by,
\begin{equation}
\xi_{i}\left(k\right)\equiv\frac{3}{2}\left[\ln\left|\frac{1}{z_{i}\left(k\right)}\right|\right]^{-1},\label{eq:Widthedge}
\end{equation}

where $\left|z_{i}\left(k\right)\right|<1$ is the smallest factor
in the eigenstate given by Eq. (\ref{eq:EdgeCompactAll-1}). We notice
that such condition is fulfilled by $z_{1}^{+}$ and $z_{2}^{-}$
between $\left(k_{l},\,k_{0}\right)$ and $\left(k_{0},\,k_{r}\right)$,
respectively, as shown in Fig. (\ref{fig: Figure3}). In Fig. (\ref{fig: Figure4})
we plot the width versus momentum for different values of $D$, where
by completeness we also have included the larger decaying factors.
In the slightly flat region around $k_{0}$ the width is almost constant.
Here, $\left|z_{1}^{+}\right|=\left|z_{2}^{+}\right|$ and $\left|z_{1}^{-}\right|=\left|z_{2}^{-}\right|$,
since the discriminant of the Eq. (\ref{eq: zCharact}) is negative
($b^{2}-4ac<0$), which gives rise a two pairs of complex conjugate
solutions, one pair with magnitude lower than one. However, out of
this complex region and between the critical values, the solutions
are real. In such region, the width associate to the solution $z_{1}^{+}$
decreases while the associate to $z_{2}^{+}$ grows quickly and diverges
at the critical point where the edge state merges into the bulk. Since
the edge state wave-function is written as a linear combination, Eq.
(\ref{eq:EdgeCompactAll-1}), then the edge state merges into the
bulk through a bifurcation of the edge state width in complete analogy
with a semi-infinite graphene \cite{Doh2013,Doh2014}.

The inset in Fig. (\ref{fig: Figure4}) shows the edge state width
as a function of $D$ at the point $k_{0}$. At such point the width
can be obtained explicitly. From the Eq. (\ref{eq: AddEdge}) for
$k=k_{0}$ we obtain $z_{1}^{+}=-z_{2}^{-}$ for $\varepsilon=3$,
and by the Eq. (\ref{eq:Widthedge}) we can write,
\begin{equation}
\xi_{0}=\frac{3}{2}\left\{ \ln\left[\frac{4\left(D/J\right)}{\sqrt{1+16\left(\frac{D}{J}\right)^{2}}-1}\right]\right\} ^{-1}.\label{eq: Widthk0}
\end{equation}

Expanding near $D=0$ the above equation takes the form,
\begin{equation}
\xi_{0}\approx-\frac{3}{2}\left[\ln\left[\frac{2D}{J}\right]\right]^{-1},\label{eq:WidhtEo}
\end{equation}
which shows that the width of the edge state vanishes logarithmically
as the DMI approaches to zero. Such behavior can be also observed
in the Fig. (\ref{fig: Figure3}), where the $z$ solutions collapses
to a single, $z=-J_{1}$. In consequence, at $D=0$ and $k=k_{0}$
the edge state is completely localized at the single line of boundary
sites as we mentioned before.

As in the fermionic model, the width of the edge state as we discussed
above could have important effects when determining the edge properties
of samples with finite size \cite{Zhou2008a,Linder2009,He2010}. For
example, it has been found that there is magnon propagation in interfaces
between ferromagnets and normal metals with a strong spin-orbit coupling,
referred to as interfacial DMI interaction \cite{Moon2013,Cho2015,Chaurasiya2016}.
It also have been proposed magnon waveguides, where the propagation
is between the interface of two topological insulators \cite{Mook2015}.
In both such cases, the width of the edge state will be useful in
the measurement of the magnon current.

\section{Conclusions}

We have derived analytical expressions for the magnon edge-state wavefunctions
and their energy spectra for a honeycomb lattice with DMI and zig-zag
edges. We demonstrate that the bosonic edge states are defined in
a region in the momentum space before merging into the bulk and the
width of the edge state can be controlled by the DMI strength. By
introducing an on-site potential at the outermost sites we found that
the magnonic ferromagnetic band structure closely resembles that of
the fermionic graphene. However, unlike graphene where the energy
spectrum is doubly degenerate, our explicit solutions demonstrate
that there is only one chiral magnon edge state at each edge as predicted
by topological approaches \cite{Owerre2016d,Owerre2016c,Kim2016a}.
Moreover, our approach is applicable for arbitrary values of the on-site
potential $\delta$ as introduced in Eq. (\ref{eq: HolePartHamil}).
The edge-magnon width, the chirality and the spin density as a function
of the DMI presented in this paper could be useful for experiments
in small sized monolayers or thin film magnets.

\section*{Acknowledgments}

We thank Wissam A. Ameen, R. Carrillo-Bastos and R. Romo for discussions.
Pierre. A. Pantale{\'o}n is sponsored by Mexico's National Council
of Science and Technology (CONACYT) under the scholarship No. 381939.

\appendix

\section{Analytical Solutions for the Hamiltonian with D=0\label{sec:Appendix}}

In this appendix, we derive the edge states for $D=0$ with an on-site
potential $\delta=1$ at the outermost sites. The coupled Harper equations
of the Hamiltonian (\ref{eq: HolePartHamil}) are given by\textbf{
\begin{eqnarray}
3\psi_{A}\left(n\right)-J_{1}\psi_{B}\left(n\right)-J_{2}\psi_{B}\left(n-1\right) & = & \varepsilon\psi_{A}\left(n\right),\nonumber \\
-J_{1}\psi_{A}\left(n\right)-J_{2}\psi_{A}\left(n+1\right)+3\psi_{B}\left(n\right) & = & \varepsilon\psi_{B}\left(n\right),\quad\label{eq: HarperBulk}
\end{eqnarray}
}where $\psi_{A,B}\left(n\right)$ are the eigenfunctions for each
sublattice. Using the Eq. (\ref{eq: MainAnzats}) we can write the
Eq. (\ref{eq: HarperBulk}) in terms of the $2-$component vector
$\phi$ as,
\begin{equation}
\left[\begin{array}{cc}
3 & -J_{1}-J_{2}z^{-1}\\
-J_{1}-J_{2}z & 3
\end{array}\right]\left[\begin{array}{c}
\phi_{A}\\
\phi_{B}
\end{array}\right]=\varepsilon\left[\begin{array}{c}
\phi_{A}\\
\phi_{B}
\end{array}\right],\label{eq: FreeMBulk}
\end{equation}
for $n>1$ (and $n<N$, for a large $N$). The non-trivial solution
gives rise to the characteristic equation,
\begin{equation}
\left(3-\varepsilon\right)^{2}-J_{1}^{2}-J_{2}^{2}-J_{1}J_{2}\left(z+z^{-1}\right)=0,\label{eq: Characzbulk}
\end{equation}

which is a quadratic equation in $z$. The edge state can be constructed
with the solution $\left|z\right|<1$ in the Eq. (\ref{eq: Characzbulk}).
As in the case for non-zero DMI, we require an additional condition
for $z$ and $\varepsilon$. From the Harper equations for the case
of $n=1$, such condition is given by,
\begin{equation}
\left(3-\varepsilon\right)^{2}-J_{1}^{2}-J_{1}J_{2}z=0.\label{eq: Charup}
\end{equation}
Hence, the existence of an edge state requires that $z$ and $\varepsilon$
must satisfied both equations (\ref{eq: Characzbulk}) and (\ref{eq: Charup}).
One such solution is $\varepsilon=3$ and $z=-J_{1}$, with eigenstate
\begin{equation}
\psi_{k}\left(n\right)=z^{n}\left[\begin{array}{c}
\phi_{A}\\
0
\end{array}\right],\label{eq: UpEigenfree}
\end{equation}
which is localized on the $A-$sublattice. This solution is consistent
with the Eq. (\ref{eq: AddEdge}) and the eigenstate given by the
Eq. (\ref{eq:EdgeCompactAll-1}) in the limit $D\rightarrow0$.

For a large $N$, we obtain the solution for the opposite edge, by
setting $n=N$ in the Harper equations and write
\begin{equation}
\left(3-\varepsilon\right)^{2}-J_{1}^{2}-J_{1}J_{2}z^{-1}=0.\label{eq: DownEdgeWav}
\end{equation}
Together with the bulk equation (\ref{eq: Characzbulk}) we find the
solutions as $\varepsilon=3$ for $z=-1/J_{1}$. Hence, the eigenstate
in terms of the ribbon size and the solution for the upper boundary,
can be written as
\begin{equation}
\psi_{k}^{\prime}\left(n\right)=z^{N-n+1}\left[\begin{array}{c}
0\\
\phi_{B}^{\prime}
\end{array}\right].\label{eq: EigenLoN}
\end{equation}
Such solution is consistent (in the limit $D\rightarrow0$) with the
eigenvector given in Eq. (\ref{eq: EigenLower}). Finally, for the
range in the momentum, since we require $\left|z\right|<1$ for the
upper and $\left|z\right|>1$ for the lower edge, the flat edge state
exists between, $k\in\left(\frac{2\pi}{3\sqrt{3}},\,\frac{4\pi}{3\sqrt{3}}\right)$
which are the coordinates of the $K$ and $K^{\prime}$ points, {[}See
Fig. \ref{fig: Figure2}(a){]}.

\bibliographystyle{apsrev4-1}

\end{document}